# Hybrid Cooperative Co-Evolution Algorithm for Deadlock-prone Distributed Assembly Flowshop Scheduling with Limited buffers Using Petri nets

Siyi Wang, Yanxiang Feng, Xiaoling Li, Guanghui Zhang, Yikang Yang

*Abstract*—**The distributed assembly flowshop scheduling problem (DAFSP) can be applied to immense manufacturing environments. In DAFSP, jobs are first processed in distributed flowshops, and then assembled into final products by an assembly machine, which usually has *limited* buffers in practical application. This limited capacity can lead to *deadlocks*, halting job completion and blocking the entire manufacturing process. However, existing scheduling methods fail to address these deadlocks in DAFSP effectively. As such, we develop a hybrid cooperative co-evolution (HCCE) algorithm for solving the deadlock-prone DAFSP by minimizing the makespan. For the first time, we use Petri nets to analyze the deadlocks in DAFSP and propose a Petri net-based deadlock amending method (IDAM), which is further integrated into HCCE to ensure the feasibility (i.e., deadlock-freeness) of solutions. Importantly, HCCE contains an elite archive (EAR) and two subpopulations. It uses the problem-specific operators for heuristic initialization and global-search. To enhance the quality and diversity of solutions, an information transfer mechanism (ITM) is developed among subpopulation and EAR, and four local-search operators are performed sequentially on each individual in EAR. Finally, comprehensive experiments demonstrate the effectiveness and superiority of the proposed HCCE algorithm.**

## I. INTRODUCTION

During the past decades, the assembly scheduling problem (ASP) has been widely used in practical manufacturing processes such as automobiles [1] and semi-conductor devices [2]. These ASPs usually comprise two stages: *manufacturing* and *assembly*. Jobs are first processed in the manufacturing stage, which could be a flowshop, jobshop, single factory, or multi-factories, etc. Then, in the assembly stage, the finished jobs will be assembled into products on a single assembly machine or parallel machines.

Due to economic globalization, enterprises have transitioned from traditional centralized production to distributed production [3]. Correspondingly, the research on the distributed ASP (DASP) attracts the attentions of many researchers worldwide. It is usually assumed that the manufacturing stage of DASP consists of distributed multi-factories and the assembly stage is a single machine. To date, many algorithms have been proposed to solve DASPs, including Genetic algorithm (GA) [4], Estimation of Distribution algorithm (EDA) [5], Iterated Greedy algorithm (IGA) [6], Variable Neighborhood Descent algorithm (VNDA) [1], etc. Their objectives are varying, such as minimizing the makespan, total completion time, or maximum lateness (one or multiple of them). Significantly, these existing literature on DASP often assume that the size of assembly buffers in the second stage is infinite [6], [7], [8]. But in real manufacturing environments, such as semiconductor manufacturing and food processing production, the capacity of assembly buffers is either nonexistent or of limited size.

The primary challenge incurred by the limited assembly buffers for DASP is the undesirable *deadlock* [9]. Precisely, if the limited assembly buffer is full of jobs but none of them can be assembled into a product, then no job can be advanced into the assembly machine and the whole system is blocked. According to [10], the deadlock of DASP is closely related to the order of finished jobs entering into the assembly buffer. To characterize and control deadlock, the whole DASP can be decoupled as two subpopulations: *job-scheduling* and *factory-scheduling*. The former determines the jobs' order entering assembly buffer, while the latter assigns each job to a specific factory. Correspondingly, the key for solving deadlock in DASP lies in the first subproblem——find a *deadlock-free* job order entering assembler buffer. Moreover, Petri net, an effective tool to model discrete event systems, have been widely used to characterize and control deadlocks [11], [12], [13], but do date no one use Petri net to analyze the deadlocks in DASP.

This paper studies a kind of deadlock-prone DASP, namely distributed assembly flowshop scheduling problem (DAFSP), with limited assembly buffers, where the manufacturing stage is a distributed *blocking flowshop* and the assembly stage is a single assembly machine. The objective is to minimize the system makespan. We develop a hybrid cooperative

This work was supported in part by the National Natural Science Foundation of China under Grant 62103062; in part by the Hebei Natural Science Foundation under Grant F2024204007; in part by the State Key Laboratory for Manufacturing Systems Engineering of Xi'an Jiaotong University under Grant sklms 2023002.

Siyi Wang, Yanxiang Feng and Yikang Yang are with the School of Automation Science and Engineering, Xi'an Jiaotong University, Xi'an 710049, China (e-mail: wangsiyi@stu.xjtu.edu.cn; fengyanxiang@xjtu.edu.cn; yangyk74@mail.xjtu.edu.cn).

Xiaoling Li is with the School of Electronics and Control Engineering, Chang'an University, Xi'an 710064, China (e-mail: xiaolingli@chd.edu.cn).

Guanghui Zhang is with the School of Information Science and Technology, Hebei Agricultural University, Baoding 071001, China, (e-mail: ghzhang@hebau.edu.cn).

**Nomenclature**

| Symbol | Definition |
|---|---|
| $u$ | Number of jobs. |
| $f$ | Number of factories. |
| $m$ | Number of machines. |
| $l$ | Number of products. |
| $i$ | Index for jobs, $i \in \{1, 2, . . . , u\}$. |
| $c$ | Index for factories, $c \in \{1, 2, . . . , f\}$. |
| $k$ | Index for machines, $k \in \{1, 2, . . . , m\}$. |
| $q$ | Index for products, $q \in \{1, 2, . . . , l\}$. |
| $\Psi$ | The capacity of assembly buffers. |
| $AP$ | The job-to-product plan. |
| $CM_{max}$ | The makespan of the manufacturing stage. |
| $CA_{max}$ | The system makespan. |
| $S_{i,k}$ | The start time of job $i$ on machine $k$. |
| $C_{i,k}$ | The completion time of job $i$ on machine $k$. |
| $SA_q$ | The start time of product $q$ on assemble machine $M_A$. |
| $CA_q$ | The completion time of product $q$ on assemble machine $M_A$. |
| $\Delta = \{\lambda, \mu\}$ | An individual of HCCE. |
| $\pi$ | A solution for DAFSP. |

co-evolution (HCCE) algorithm to solve these deadlock-prone DAFSPs. First, an *assembly procedure Petri net* (APP) is established to model the entering of jobs into the assembly buffer. Based on banker's algorithm (BA), a Petri net-based deadlock amending method (IDAM) is proposed to ensure the feasibility or deadlock-freeness of the jobs' order. Then, IDAM is embeded into HCCE to coordinate deadlock control and scheduling. Importantly, HCCE employs the modified cooperative co-evolution algorithm (mCCEA) framework, including two subpopulations and an elite archive (EAR). These subpopulations primarily consist of job-permutations and factory-permutations, respectively, corresponding to the subpopulations of job-scheduling and factory-scheduling. The main contributions are summarized as follows.

1. For the first time we use Petri nets for analyzing deadlocks in DAFSP and a strategy IDAM with polynomial complexity is proposed.
2. We incorporate IDAM into HCCE to ensure the feasibility or deadlock-freeness of the solution of DAFSP, and the makespan of a solution is calculated by a novel backward method which maintains the deadlock-freeness of solution.
3. Two problem-specific heuristic operators are constructed for initialization and global evolution of proposed HCCE algorithm.
4. To improve the quality and diversity of individuals, an ITM is used to exchange information among subpopulations and EAR. Meanwhile, four local search operators are applied sequentially for each individual in EAR, enhancing the searchability of the algorithm.

To evaluate the performance of HCCE in solving the DAFSP, three variants of HCCE and three state-of-the-art metaheuristic algorithms HHMA [7], EDMBO [8], and PBIGA [6] are selected for comparison. The experimental results demonstrate that: 1) HCCE outperforms its three variants in solving various instances, verifying the effectiveness of the specially designed components of HCCE. 2) HCCE outperforms the three compared metaheuristic algorithms, verifying the effectiveness of HCCE in solving deadlock-prone DAFSP.

The rest of this article is organized as follows. Section II reviews the related literature. Section III describes the DAFSP and the methods to check and amend the deadlock. Section IV presents the HCCE algorithm. Section V presents the results of computational experiments. Finally, Section VI summarizes our work and suggests directions for future research.

## II. Literature Review

Extensive and systematic studies have been conducted on ASP [14]. Lee *et al.* [15] first consider the three-machine assembly flowshop scheduling problem to minimize makespan, where two machines are used in the first stage and the other in the assembly stage. The work [16] considers multiple parallel machines at the job processing stage and develops a heuristic algorithm with a worst-case ratio bound to minimize the makespan. Furthermore, the ASP with multiple non-identical assembly machines is studied in [16], where a hybrid algorithm is presented combining the variable neighborhood search with a heuristic.

The DASP has been investigated in recent years. This problem involves a manufacturing stage comprising several identical production factories. Hatami *et al.* [17] study this DASP to minimize the makespan by presenting a mixed integer linear programming model and a VND. Several meta-heuristic algorithms are developed to improve the scheduling efficiency [14]. For example, an effective Hyper Heuristic-based Memetic Algorithm (HHMA) [7] and an Estimation of Distribution Algorithm-based Memetic Algorithm (EDAMA) [5] are presented for DAPFSP for minimizing makespan. An IGA is used in [6] and [18] based on groupthink to minimize the total flow time. Considering the blocking constraint in the manufacturing stage, an effective Water Wave Optimization Algorithm with problem-specific knowledge (KWWO) is presented for solving the distributed assembly blocking flow-shop scheduling problem (DABFSP). The above articles all focus on minimizing a single objective. However, He *et al.* [19] minimize makespan, total flow time, and total energy consumption simultaneously by a Greedy Cooperative Co-Evolutionary Algorithm (CCEA). Additionally, Wang and Wang [20] minimize the total tardiness and energy consumption simultaneously by a cooperative memetic algorithm with feedback.

The CCEA is first presented by Potter and De Jong [21]. It has been used to solve many optimization problems, including power systems [22], [23], vehicle routing problems [24], [25], satellite-module layout [26], etc. Recently, a CCEA with a new cooperation mechanism, known as reference sharing, is proposed for function optimization problems [27]. Lei [28] provides a co-evolutionary genetic algorithm (CGA) for a fuzzy flexible job shop scheduling problem. Zheng and Wang [29] propose a CCEA for a resource-constrained unrelated parallel machine green scheduling problem for minimizing the makespan and the total carbon emission. Besides, Pan [30] develops a co-evolutionary artificial bee colony (CCABC) algorithm for a steelmaking-continuous casting scheduling problem to minimize the makespan and charge waiting times.

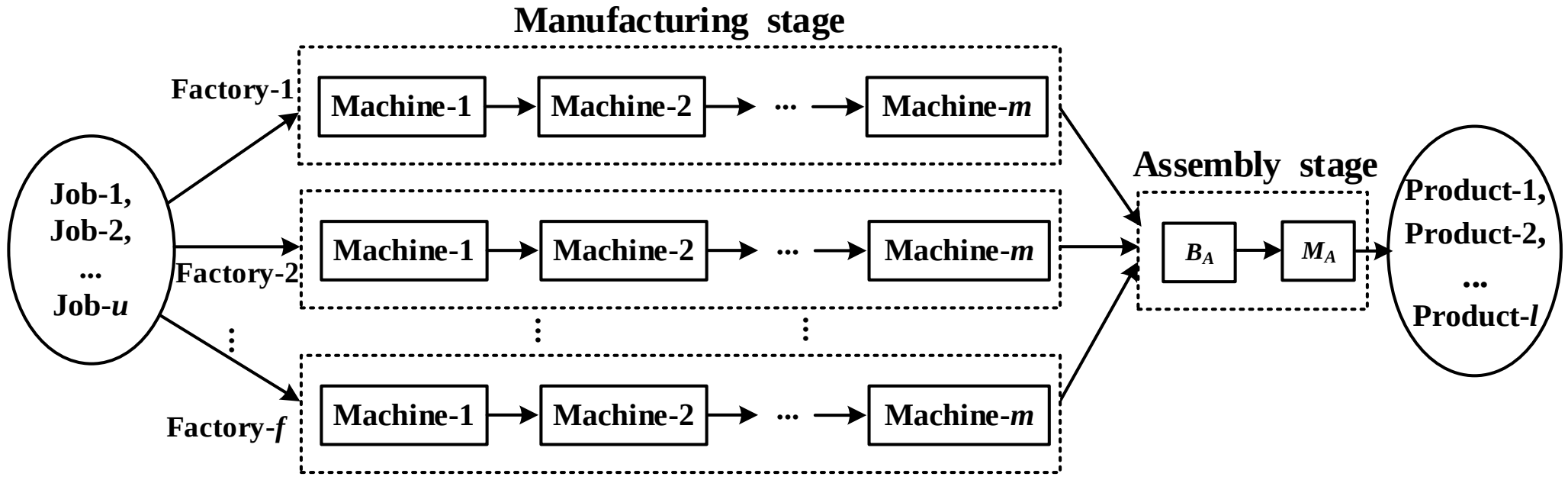

Fig. 1. The diagram of a DAFSP.

Although deadlocks are a crucial aspect of manufacturing system evolution, as far as we know, there is no literature resolving the deadlock issue in DAFSP. The recent studies for solving deadlocks mainly focus on automated manufacturing systems (AMSs). Petri net is an important tool to model and characterize deadlocks [31]. Liu *et al*. [32] develop a transition cover-based design of Petri net deadlock controllers for AMSs. The robust deadlock control problem for AMSs with unreliable resources is studied in [33], [34] where Petri nets model the unreliable systems. Feng *et al*. [35] focus on the robust deadlock prevention problem for AMSs with a type of unreliable resources. Considering the assembly operation, Wu *et al*. [36] study the deadlock problem and model this system using resource-oriented Petri nets. Hu and Zhou [37] use a mathematical programming method to derive each deadlock in an iterative way, and synthesize a live controlled net.

## III. DEADLOCK-PRONE DAFSP

### A. Problem description and modeling

As shown in Fig. 1, the studied DAFSP has $f$ identical factories $F = \{c_1, c_2,\ldots, c_f\}$ and an assembly machine $M_A$. It aims to produce $l$ products $L = \{q_1, q_2,\ldots, q_l\}$ by scheduling $u$ jobs $U = \{i_1, i_2,\ldots, i_u\}$. Each factory is a flow shop with $m$ machines $M = \{k_1, k_2,\ldots, k_m\}$. The whole DAFSP consists of two stages: *manufacturing* and *assembly*. In the first manufacturing stage, all factories process jobs simultaneously, the finished jobs will be removed into the *assembly buffers* $B_A$ with the capacity $\Psi$. The finished jobs in $B_A$ will be assembled into products by machine $M_A$ according to the *job-to-product plan* $AP = \{AP_q \mid q = 1, 2,\ldots, l\}$, where product $q \in L$ is assembled by $|AP_q|$ jobs in $AP_q$. Only after all jobs in $AP_q$ are removed into $B_A$, the assembling of product $q$ starts.

The *makespan* of the manufacturing (resp. assembly) stage is denoted by $CM_{max}$ (resp. $CA_{max}$), which equals to the maximum completion time of all jobs (resp. products). Their calculation is given in Section III.D. Herein, $CA_{max}$ is also called *system makespan*, and the objective of this paper is to minimize $CA_{max}$. We make the following assumptions for DAFSP:

(1) All jobs and machines are available at time zero. The processing time of job $i$ on machine $k$ is predefined as $p_{i,k}$, and the assembly time of product $q$ on machine $M_A$ is set as $pA_q$.

(2) In each factory, all jobs have to follow the same path on every machine, and each job must path through all the machines within its assigned factory.

(3) A job can only be processed on a single machine at any time, and each machine can process one job at a time.

(4) The setup time is included in the processing time and transportation time of jobs are disregarded.

(5) Each job can only belong to one product plan, and the finished job will be removed into $B_A$ immediately if $B_A$ is available.

(6) Each job can be processed at any time after its previous job is finished, without the necessity of immediate processing.

Moreover, we introduce the following variables to describe DAFSP.

- $S_{i,k}$: start time of job $i$ on machine $k$;
- $C_{i,k}$: completion time of job $i$ on machine $k$;
- $SA_q$: start time of product $q$ on assembly machine $M_A$;
- $CA_q$: completion time of product $q$ on assembly machine $M_A$;

The completion time of job $i$ on the last machine-$m$ is represented by $C_{i,m}$. Since only after all jobs in $AP_q$ are manufactured, product $q$ can be assembled, the following inequality holds.

$$SA_q \geq \max_{i\in AP_q}\{C_{i,m}\} \tag{1}$$

### B. Solution coding and deadlock

The whole DAFSP can be divided into two subproblems: *job scheduling* and *factory scheduling*. The former subproblem determines the processing order of jobs entering into buffer $B_A$. The latter subproblem determines the assignment of jobs to factories. Thus, an individual for DAFSP is coded as $\Delta = \{\lambda, \mu\}$, which consists of two permutations $\lambda = (\lambda[1], \lambda[2],\ldots, \lambda[u])$ and $\mu = (\mu[1], \mu[2],\ldots, \mu[u])$ denoting the permutations of jobs and factories, respectively. Note that each job appears exactly once in $\lambda$.

For a coding $\Delta = \{\lambda, \mu\}$, after allocating job $i$ to the factory $\mu[i]$, and sorting all jobs assigned to each factory according to their order in $\lambda$, we obtain a unique *solution* $\boldsymbol{\pi} = \{\pi_c \mid c \in F\}$ of DAFSP. For example, as illustrated by Fig.2, for $\Delta = \{\lambda, \mu\}$ where $\lambda = (i_1, i_4, i_3, i_2, i_5)$ and $\mu = (c_2, c_2, c_1, c_1, c_2)$, the solution extracted from Δ is $\boldsymbol{\pi} = \{\pi_1, \pi_2\}$, where $\pi_1 = \{i_4, i_3\}$ and $\pi_2 = \{i_1, i_2, i_5\}$.

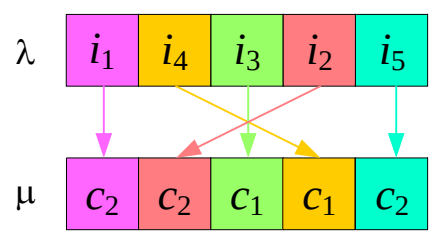

Fig. 2. An example of a solution coding Δ = {λ, μ}.

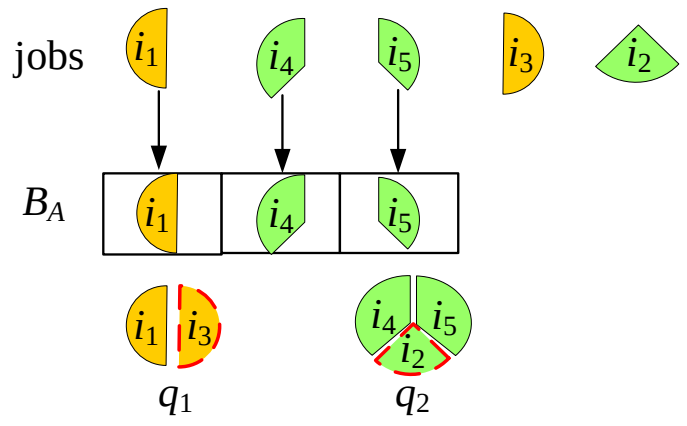

Fig. 3. The example of deadlock.

If $B_A$ is full of jobs but none of them can be removed into $M_A$ for assembling, then a blockage, namely *deadlock*, is resulted.

*Example 1*: Consider a DAFSP with five jobs $i_1$–$i_5$ and two products $q_1$ and $q_2$ with $AP_1 = \{i_1, i_3\}$ and $AP_2 = \{i_2, i_4, i_5\}$. The capacity of $B_A$ is $\Psi = 3$. As shown in Fig. 3, according to $\lambda = (i_1, i_4, i_5, i_3, i_2)$, assuming that after $i_1$, $i_4$, and $i_5$ complete the processing and enter $B_A$, the buffer becomes saturated. Due to the lack of job $i_3$ (resp. $i_2$), product $q_1$ (resp. $q_2$) cannot be produced by assembling jobs in $B_A$. Thus, jobs $i_2$ and $i_3$ cannot be removed into $B_A$ anymore and the system is trapped in *deadlock*.

As illustrated by Example 1, deadlock is closely related to the order of the jobs' arrival at buffer $B_A$, i.e., the permutation λ. Therefore, it is imperative to convert λ into a deadlock-free one. The next subsection will develop a deadlock-amending method jointly using Petri net model and Banker's algorithm (BA).

### C. Deadlock-amending method based on Petri net and BA

#### 1) Petri net model

This subsection develops an *assembly procedure Petri net* (APP) for modeling the entering of jobs into $B_A$. See Appendix A for the basic definitions of Petri net.

*Definition 1*: The APP $(N, M_0)$ model of a DAFSP is constructed by the following steps:

Step 1: Establish a path $t_q p_q^e$ for each product $q \in L$, as well as a path $p_i t_i p_i^e$ for each job $i \in AP_q$, and add an arc $(p_i^e, t_q)$ between every $p_i^e$ and $t_q$.

Step 2: Let $p_B$ be a place representing the buffer $B_A$. Add arcs $(p_B, t_i)$ (resp. $(t_q, p_B)$) for each transition $t_i$ (resp. $t_q$) with weight 1 (resp. $|AP_j|$).

Step 3: Let $M_0$ be the initial marking where $M_0(p_i) = 1$, $M_0(p_i^e) = 0$, $\forall i \in U$, $M_0(p_B) = \Psi$, and $M_0(p_q^e) = 0$, $\forall q \in L$.

*Example 2:* Consider the DAFSP in Fig. 3 with five jobs $i_1$–$i_5$ and two products $q_1$ and $q_2$ where $AP_1 = \{i_1, i_3\}$ and $AP_2 = \{i_2, i_4, i_5\}$. The capacity of buffer $B_A$ is $\Psi = 3$. Fig. 4(a) illustrates the APP model $(N, M_0)$.

*Remark 1:* In $(N, M_0)$, place $p_i$ with a token represents that job $i$ is finished on the last machine $m$, the firing of $t_i$ represents the moving of job $i$ into $B_A$, while $p_i^e$ with a token denotes the waiting of job $i$ in $B_A$. The firing of $t_q$ represents the start of the assembly operation of product $q$, and $p_q^e$ denotes the finished product $q$. Buffers in $p_B$ can be used or released by firing transitions $t_i$ or $t_q$. Initially, $M_0$ implies that all jobs complete their processing in the first stage and all buffers of $B_A$ are available. When all jobs are assembled and products are produced, each $p_q^e$ is marked. Then APP reaches a *final state* $M_E$, where $M_E(p_i) = M_E(p_i^e) = 0$, $\forall i \in U$, $M_0(p_B) = \Psi$, and $M_E(p_q^e) = 1$, $\forall q \in L$.

*Remark 2:* We assume that the assembly transition $t_q$, $q \in L$, is fired under a state $M$ as long as it is enabled, i.e., $\forall p_i^e \in {}^\bullet t_q$, $M(p) > 0$. This is, after all jobs in $AP_q$ are removed into $B_A$, the assembly of product $q$ starts immediately, releasing the occupied buffers.

Given a permutation λ, by converting each job $i$ of λ into transition $t_i$, we can extract a transition sequence $\alpha(\lambda)$, representing the order of jobs entering buffer $B_A$. According to the firing sequence of $\alpha(\lambda)$, the system may enter a deadlock.

*Example 3:* Consider the APP model $(N, M_0)$ in Fig. 4(a). For a permutation $\lambda = (i_1, i_4, i_5, i_3, i_2)$, its corresponding transition sequence is $\alpha(\lambda) = t_{i1}t_{i4}t_{i5}t_{i3}t_{i2}$. However, only firing the first three transitions $t_{i1}t_{i4}t_{i5}$, a deadlock $M = p_{i2} + p_{i4}^e + p_{i5}^e + p_{i1}^e + p_{i3}$ is resulted, shown in Fig. 4(b). Particularly, at $M$, place $p_B$ is empty, so transition $t_{i2}$ or $t_{i3}$ cannot fire and no token is flowed into $p_{i2}^e$ or $p_{i3}^e$. Then, transitions $t_{q1}$ and $t_{q2}$ cannot fire and no occupied buffer $B_A$ is released, i.e., $p_B$ cannot be marked forever. Thus, $M$ is a deadlock.

#### 2) Improved BA to check the safeness of a marking

According to the supervisory control theory in DEDS [38], a state $M$ of APP $(N, M_0)$ is *safe* if it can reach the final state $M_E$. That is, the *safeness* of state $M$ implies that all finished jobs can be assembled into products starting from $M$. We propose an improved BA (IBA) to determine the safeness of a specific marking.

**Algorithm 1** IBA (Improved Bank's Algorithm).

**Input:** an APP model $(N, M_0)$ and a marking $M$
**Output:** True or False
$Flag_w = 0$; /* $Flag_w = 1$ denotes that the safety detection result of the input marking $M$ is obtained; otherwise, $Flag_w = 0$.*/
Let $\Omega = \{q \in L \mid M(p_q^e) = 0\}$
Let $\Theta = [\Theta_1, \Theta_2, \ldots, \Theta_l]$ where $\Theta_q = |AP_q| - \sum_{i \in APq} M(p_i^e)$;
Let $M_{cu} = M$;
**While**(!$Flag_w$)
  **if** $\Omega = \varnothing$
    $Flag_{df}$ = True, $Flag_w = 1$; /* $Flag_{df}$ = True denotes the input marking is deadlock-free; otherwise, $Flag_{df}$ = False.*/
  **else**
    Let $\Pi = \{q \in L \mid M_{cu}(p_B) \geq \Theta_q\}$
    **If** $\Pi \neq \varnothing$
      Select $q \in \Pi$;
      $\Omega = \Omega \setminus \{q\}$;
      Let $M_1 = M_{cu}$;
      Set $M_1(p_q^e) = 1$, $M_1(p_B) = M_1(p_B) + |AP_q| - \Theta_q$,

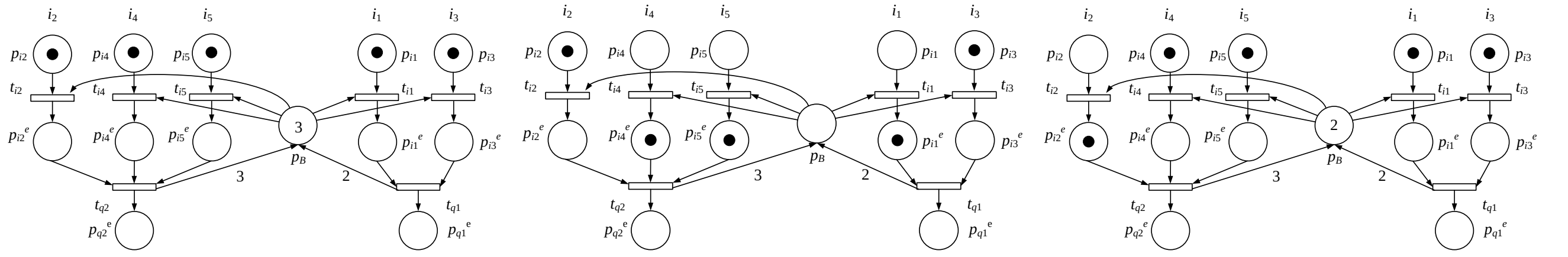

(a) Initial marking $M_0$ in ($N$, $M_0$) (b) Deadlock $M$ in ($N$, $M_0$) (c) Safe state $M'$ in ($N$, $M_0$)

Fig. 4. APP model of DAFSP.

$M_1(p_i^e)=M_1(p_i)= 0\ \forall i\in AP_q$;
$M_{cu} = M_1$.
**else**
$Flag_{df}$ = false, $Flag_w = 1$;
**end**
**end**
**end**
**return** $Flag_{df}$;

In IBA, first let $\Omega = \{q \in L \mid M(p_q^e) = 0\}$ collect all the non-finished products under $M$, and a sequence $\Theta = [\Theta_1, \Theta_2, \ldots, \Theta_l]$ with $\Theta_q = |AP_q| - \sum_{i\in APq}M(p_i^e)$ denote the number of jobs in $AP_q$ that are not removed into $B_A$. The currently-analyzing marking $M_{cu}$ is initially set as the input $M$. Then a loop is executed, which contains two main parts:

i) $\Omega = \varnothing$, all products are assembled, $M_{cu} = M_E$, and hence $M$ is safe, True is returned.

ii) $\Omega \neq \varnothing$, we set $\Pi = \{q \in L \mid M_{cu}(p_B) \geq \Theta_q\}$. If $\Pi \neq \varnothing$, we select a product $q \in \Pi$, assemble it under $M_{cu}$, obtain a new marking $M_1$ by emptying all places $p_i$ and $p_i^e$, $\forall i \in AP_q$, and adding a token into $p_q^e$. Then, update $M_{cu} = M_1$. Otherwise, if $\Pi = \varnothing$, it means there are no adequate buffers to support the assembly of the remaining products, hence False is returned.

*Example 4:* Consider the APP in Fig. 4 (a) and the state $M$ in Fig. 4 (b). By IBA, we have $\Omega = \{q_1, q_2\}$, $\Theta = [1, 1]$, $M_{cu}(p_B) = 0$, $\Omega \neq \varnothing$, and $\Pi = \varnothing$, indicating that products $q_1$ and $q_2$ can never be assembled under $M$. For another state $M' = p_1 + p_2^e + p_3 + p_4 + p_5 + 2p_B$ shown Fig. 4(c), we have $\Omega = \{q_1, q_2\}$, $\Theta = [2, 2]$, and $M_{cu}(p_B) = 2$. First, select $q_1$ from $\Pi$, since there are adequate buffers to store the jobs for assembling $q_1$, i.e., jobs in $p_{i1}$ and $p_{i3}$, we empty places $p_{i1}$, $p_{i3}$, $p_{i1}^e$ and $p_{i3}^e$, and a token is added into $p_{q1}^e$, resulting a new state $M_1 = p_2^e + p_4 + p_5 + p_{q1}^e + 2p_B$. Similarly, product $q_2$ can be assembled as well under $M_1$. Therefore, $M'$ is safe and the output is True.

3) *IBA-based deadlock-amending method*

Given an APP ($N$, $M_0$) and a permutation $\lambda$, by the aid of Algorithm IBA, we develop a polynomial-complexity algorithm, namely IDAM, to determine whether $\lambda$ is deadlock-free, and if not, we convert $\lambda$ into a deadlock-free permutation.

**Algorithm 2** IDAM (IBA-based deadlock-amending method).

**Input:** an APP model ($N$, $M_0$) and a $\lambda$-sequence
**Output:** a deadlock-free sequence $\lambda'$.
1: Obtain $\alpha(\lambda) = t_1t_2\ldots t_n$, i.e., the transition sequence corresponding to $\lambda$.
2: Let $M_{cu} = M_0$;
3: $\gamma = \alpha(\lambda)$;
4: **for each** $r \in [1, |\alpha(\lambda)|]$
5: $Flag_t = 0$ ;/* $Flag_t = 0$ denotes that safe transition is not found; otherwise, $Flag_t = 1$.
6: **while**(!$Flag_t$)
7: Let $t$ be the $r$-th transition of $\gamma$;
8: $M_{cu}[t> M$;
9: **if** IBA($M$) = True
10: $M_{cu} = M$;
11: $Flag_t = 1$;
12: **else**
13: remove $t$ to the end of $\gamma$;
14: **end**
15: **end**
16: **end**
17: Extract a job permutation $\lambda'$ from $\gamma$;
18: Output $\lambda'$.

Algorithm IDAM first obtains the transition sequence $\alpha(\lambda)$ from $\lambda$, and set the currently-analyzing $M_{cu} = M_0$ and $\gamma = \alpha(\lambda)$. Then a loop **while** (Lines 5–14) is performed $|\alpha(\lambda)|$ times. Specifically, during the $r$-th time of executing the loop, let $t$ denote the $r$-th transition of $\gamma$, $M$ be the state after firing $t$ at $M_{cu}$, i.e., $M_{cu}[t> M$, if IBA($M$) = True, $M$ is safe, then let $M_{cu} = M$, and this loop ends; otherwise, remove $t$ to the end of $\gamma$ and the current loop continues.

*Remark 3:* In APP model ($N$, $M_0$), each transition is fired exactly once before reaching the final marking $M_E$. Given a permutation $\lambda$ and its corresponding transition sequence $\alpha(\lambda)$, algorithm IDAM can find a feasible transition sequence $\gamma$, i.e., an updated version of $\alpha(\lambda)$, according to which $M_E$ can be reached from $M_0$. That means, all products can be assembled successfully by firing the transitions in $\gamma$ sequentially. Therefore, the job permutation $\lambda'$ extracted from $\gamma$ is deadlock-free.

*Computational complexity of Algorithm IDAM:* The entire IDAM method repeats $|\alpha(\lambda)| = u$ iterations, where $u$ is the number of jobs. In the $r$-th iteration, at most $(u - r + 1)$ transitions are checked. Thus, the complexity of IDAM is $\mathcal{O}(u^2)$, i.e., IDAM is of polynomial complexity.

*Example 5:* Consider the APP in Fig. 4(a) and a permutation $\lambda = (i_1, i_4, i_5, i_3, i_2)$, $\alpha(\lambda) = t_{i1}t_{i4}t_{i5}t_{i3}t_{i2}$. The procedure of implementing IDAM on $\lambda$ is illustrated in Fig. 5. First let $\gamma = \alpha(\lambda)$. Then as shown in Fig. 5(a) (resp. (b)), after firing the first (resp. second) transition $t_{i1}$ (resp. $t_{i4}$) of $\gamma$, the obtained state is determined safe by IBA. However, the resulted state after firing the third transition $t_{i5}$ is detected as False by IBA. So, $t_{i5}$ needs to be remove to the end of γ, generating a new sequence γ = $t_{i1}t_{i4}t_{i3}t_{i2}t_{i5}$ in Fig. 5(c). Subsequently, we continue analyzing the third transition $t_{i3}$ of γ, and the resulting state, shown in Fig. 5(d), is checked safe by IBA. Similarly, by firing the remaining transition $t_{i2}$ and $t_{i5}$, we can find that the resulted states in Fig. 5(e) and (f) are safe. Thus, the final marking $M_E$ can be reached from $M_0$ by firing the transition in γ sequentially. That is, the updated sequence γ is feasible, and the corresponding processing sequence λ′ = ($i_1$, $i_4$, $i_3$, $i_2$, $i_5$) is deadlock-free.

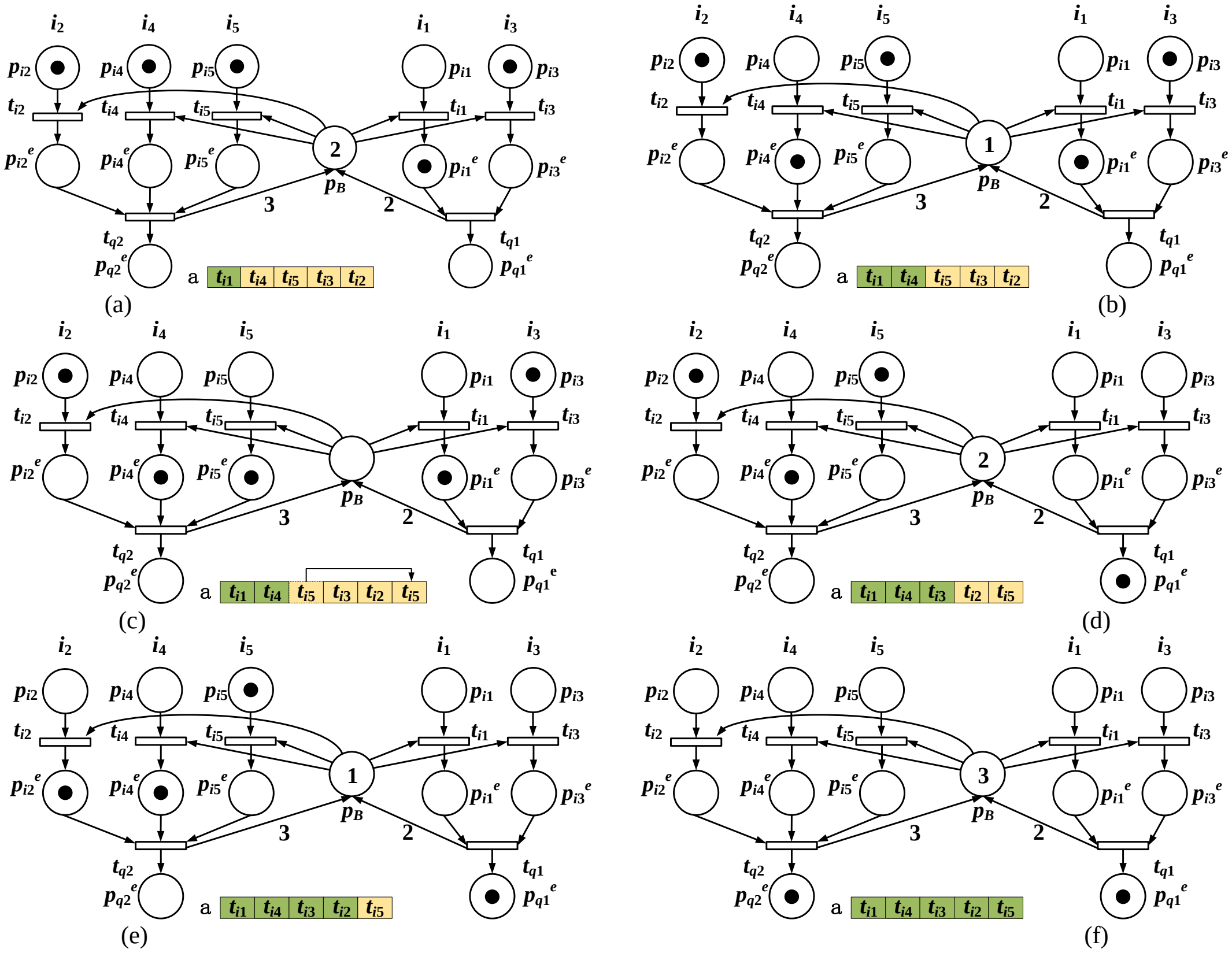

Fig.5. Processing of IDAM. (a) $t_{i1}$ is fired. (b) $t_{i4}$ is fired. (c) $t_{i5}$ is moved to the end of α. (d) $t_{i3}$ is fired. (e) $t_{i2}$ is fired. (f) $t_{i5}$ is fired.

### *D. Makespan calculation*

Given a coding Δ = {λ, μ}, we first implement algorithm IDAM to amend λ, obtaining a deadlock-free λ′. Then, a deadlock-free solution **π** can be obtained from Δ′ = {λ′, μ}. The system makespan of **π** is computed in a backward way, so as to maintain the deadlock-freeness of **π**.

First let σ be the sequence of assembling products extracted from λ′. The makespans $CM_{max}$ and $CA_{max}$ of **π** = {$\pi_c$ | $c$ ∈ $F$} can be computed based λ′ and σ according to equations (2)–(9), where $\pi_c$, λ′, and σ can be written as $\pi_c = \{i_{c(1)}, i_{c(2)},\ldots, i_{c(|\pi_c|)}\}$, λ′ = (λ′(1), λ′(2),…, λ′($u$)), and σ = (σ(1), σ(2),…, σ($l$)) for convenience, respectively.

$$\begin{cases} C_{\lambda'(u),m} = Time \\ S_{\lambda'(u),m} = C_{\lambda'(u),m} - p_{\lambda'(u),m} \end{cases} \tag{2}$$

$$\begin{cases} C_{\lambda'(h),m} = \begin{cases} S_{\lambda'(h+1),m} & \text{if } \lambda'(h),\lambda'(h+1) \in \pi_c, \\ & \forall h \in U/\{u\}, \forall c \in F \\ C_{\lambda'(h+1),m} - 1 & \text{otherwise} \end{cases} \\ S_{\lambda'(h),m} = C_{\lambda'(h),m} - p_{\lambda'(h),m} \end{cases} \tag{3}$$

$$\begin{cases} C_{i_{c(|\pi_c|)},k} = S_{i_{c(|\pi_c|)},k+1} \\ S_{i_{c(|\pi_c|)},k} = C_{i_{c(|\pi_c|)},k} - p_{i_{c(|\pi_c|)},k} \end{cases} \quad \forall c \in F, \forall k \in M/\{m\} \tag{4}$$

$$\begin{cases} C_{i_{c(h)},k} = \min\left(S_{i_{c(h)},k+1}, S_{i_{c(h+1)},k}\right) \\ S_{i_{c(h)},k} = C_{i_{c(h)},k} - p_{i_{c(h)},k} \end{cases}$$
$$\forall h \in \{1,2,\ldots,|\pi_c|-1\}, \forall c \in F, \forall k \in M/\{m\} \tag{5}$$

$$CM_{\max} = C_{\lambda'(u),m} - \min_{i \in U}\left(S_{i,1}\right) \tag{6}$$

$$\begin{cases} SA_{\sigma(1)} = \max_{i \in AP_{\sigma(1)}}\{C_{i,m}\} \\ CA_{\sigma(1)} = SA_{\sigma(1)} + pA_{\sigma(1)} \end{cases} \tag{7}$$

$$\begin{cases} SA_{\sigma(j)} = \max\left\{CA_{\sigma(j-1)}, \max_{i \in AP_{\sigma(j)}}\{C_{i,m}\}\right\} \\ CA_{\sigma(j)} = SA_{\sigma(j)} + pA_{\sigma(j)} \end{cases} \quad \forall j \in L/\{1\} \tag{8}$$

$$CA_{max} = \max_{j \in L} CA_{\sigma(j)} - \min_{i \in U}\left(S_{i,1}\right) \tag{9}$$

Table I
THE PROCESSING TIME OF JOBS $i_1$– $i_5$ ON MACHINES

| | $k_1$ | $k_2$ | $k_3$ |
|---|---|---|---|
| $i_1$ | 5 | 3 | 6 |
| $i_2$ | 4 | 3 | 5 |
| $i_3$ | 3 | 3 | 4 |
| $i_4$ | 6 | 4 | 6 |
| $i_5$ | 4 | 6 | 4 |

Table II
THE PROCESSING TIME OF PRODUCTS $q_1$ - $q_2$ ON MACHINES

| | $M_A$ |
|---|---|
| $q_1$ | 4 |
| $q_2$ | 5 |

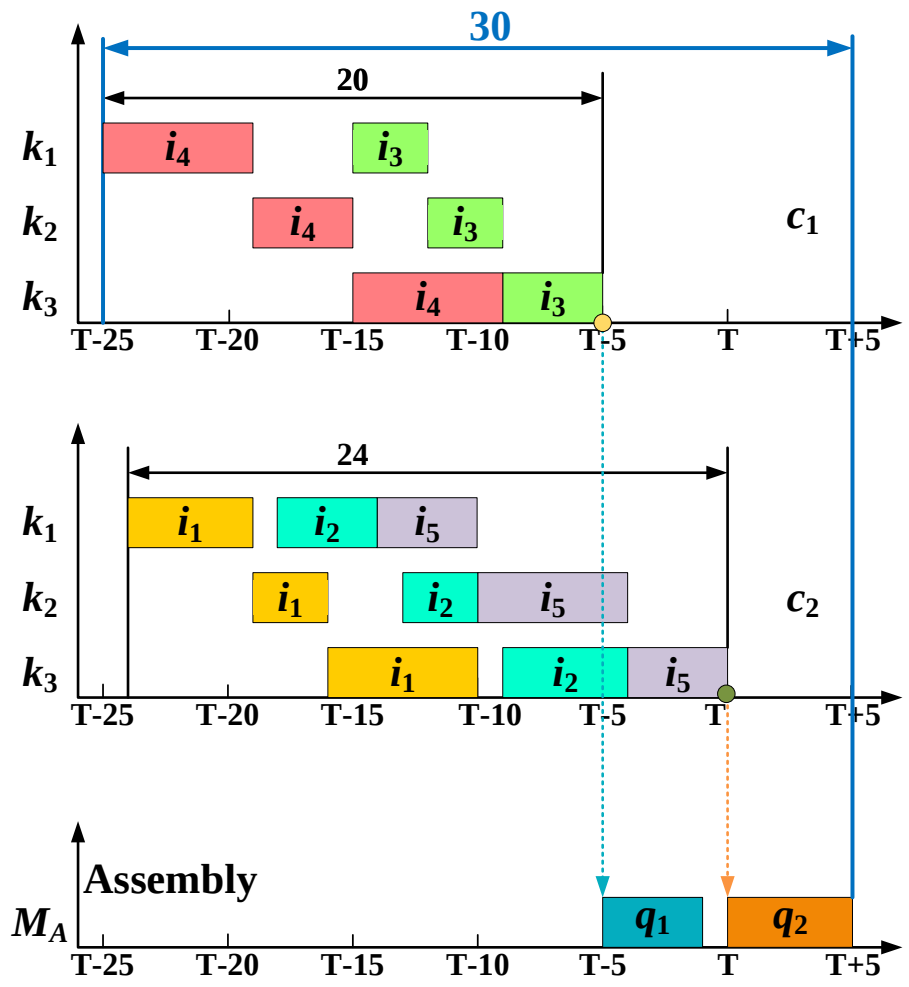

Fig. 6. Gantt chart of a solution of DAFSP.

In equation (2), let the scheduling time *Time* be the completion time of the last job $\lambda'(u)$ on the last machine $m$ in the manufacturing stage, i.e., $C_{\lambda'(u), m} = Time$, and its start time can be computed. Equation (3) determines $C_{i, m}$ and $S_{i, m}$ for each job $i \in \{\lambda'(1), \lambda'(2),\ldots, \lambda'(u-1)\}$ on machine $m$. In particular, if two consecutive jobs $\lambda'(h)$ and $\lambda'(h+1)$ are assigned to the same factory, we set $C_{\lambda'(h), m} = S_{\lambda'(h+1), m}$. Otherwise, to ensure $\lambda'(h)$ is finished before $\lambda'(h+1)$, we set $C_{\lambda'(h), m} = C_{\lambda'(h+1), m} - 1$. Equations (4) and (5) compute the completion time and start time for the last job $i_{c(|\pi c|)}$ and the remaining jobs of $\pi_c$ on every machine $k \in M / \{m\}$, respectively. Equation (6) obtains the makespan of the manufacturing stage. Equations (7) and (8) obtain the start time and completion time of the first product $\sigma(1)$ and $\sigma(j)$, $j > 1$, respectively. Equation (9) computes the system makespan $CA_{max}$.

*Example 6:* Reconsider the DAFSP in Example 2, its APP model is shown in Fig. 4(a). The processing time of jobs $i_1$–$i_5$ and the products $q_1$ and $q_2$ on various machines are illustrated in Table I and II, respectively. For $\Delta = \{\lambda, \mu\}$, where $\lambda = (i_1, i_4, i_5, i_3, i_2)$ and $\mu = (c_2, c_2, c_1, c_1, c_2)$, first $\lambda$ is converted into a deadlock-free $\lambda' = (i_1, i_4, i_3, i_2, i_5)$ by Algorithm IDAM. Thus, the solution extracted from $\{\lambda', \mu\}$ is $\boldsymbol{\pi} = \{\pi_1, \pi_2\}$, where $\pi_1 = \{i_4, i_3\}$ and $\pi_2 = \{i_1, i_2, i_5\}$. Meanwhile, $\sigma = (q_1, q_2)$ is obtained from $\lambda'$. According to (2)–(9), we obtain $CM_{max} = 25$, $CA_{max} = 30$, and the corresponding Gantt chart is provided in Fig. 6, where the scheduling time *Time* is denoted as symbol "T" for brevity

## IV. HCCE ALGORITHM FOR DAFSP

The proposed Hybrid Cooperative Co-Evolution (HCCE) algorithm for DAFSP is introduced in this section. The *individual* of HCCE is referred to the double-permutation coding $\Delta = \{\lambda, \mu\}$. The main components of HCCE are: *m*CCEA framework, initialization method, global search, information transfer mechanism (ITM), local search methods, and reinitialization strategy.

### A. mCCEA framework

The proposed HCCE includes two subpopulations, namely $\Pi_1$ and $\Pi_2$, which are composed of $\lambda$ permutations and $\mu$ permutations. Let $\Pi_1$ and $\Pi_2$ have the same size, denoted as *PS*.

Similar to the work [40], each entity in $\Pi_1$ is represented by $<\lambda^n, col_1[n]>$, where $n = 1, 2,\ldots, PS$. The integer $col_1[n]$, ranging from 1 to *PS*, is called *collaborator* of $\lambda^n$, which indicates that $\lambda^n$ is associated with the $col_1[n]$-th permutation $\mu^{col1[n]}$ in $\Pi_2$. Then, $\lambda^n$ and its collaborator constitute an individual $\{\lambda^n, \mu^{col1[n]}\}$. Similarly, each entity in $\Pi_2$ is represented by $<\mu^n, col_2[n]>$, where $col_2[n]$ is an integer from 1 to *PS*, representing the collaborator index in $\Pi_1$.

At most 2×*PS* individuals can be obtained by combining the $\lambda$ permutations and $\mu$ permutations with their collaborators. We use EAR, denoted by $\Phi$, to store the superior *ep*% individuals, i.e., $AS = ep\% \times PS$ is the size of $\Phi$.

For example, consider the *m*CCEA population shown in Fig. 7, where $PS = 3$ and $AS = 1$. The entity $<\lambda^1, 3>$ in $\Pi_1$ indicates that the collaborator of $\lambda^1$ is $\mu^3$, and a new individual $\Delta^1 = \{\lambda^1, \mu^3\}$ can be obtained. Similarly, for the entity $<\mu^1, 1> \in \Pi_2$, the collaborator of $\mu^1$ is $\lambda^1$, and the corresponding individual is $\Delta^2 = \{\lambda^1, \mu^1\}$. Following this scheme, another four individuals $\Delta^3 = \{\lambda^2, \mu^1\}$, $\Delta^4 = \{\lambda^2, \mu^2\}$, $\Delta^5 = \{\lambda^3, \mu^2\}$, and $\Delta^6 = \{\lambda^3, \mu^3\}$ are obtained. Among them, assuming that $\Delta^1$ is the best individual, therefore, it can be used to constitute the EAR $\Phi$.

### B. Initialization

1) *Heuristic initialization methods*

Let $\vartheta_i = \sum^{m}_{k=1} p_{i,k}$ be the sum of processing time for a job $i$, and $\sum_{i \in APq} \vartheta_i$ be the sum of processing time of jobs that are used for assembling product $q$. We first present two problem-specific heuristic operators $H_1$ and $H_2$.

$H_1$*:* Given a job permutation $\lambda = (\lambda[1], \lambda[2],\ldots, \lambda[u])$, we assign each job $i_n = \lambda[n]$, $n \in \{1,2,\ldots, u\}$, to all possible factories and calculate the corresponding makespan $CM_{max}$, and the factory with the lowest $CM_{max}$ is selected to process job $i_n$. Finally, a complete factory permutation $\mu$ is obtained.

$H_2$: Given a factory permutation $\mu$, generate a job permutation $\gamma = (\gamma[1], \gamma[2],\ldots, \gamma[u])$ by sorting all jobs in descending order of the sum of processing time. Let the job permutation $\lambda$ be an empty sequence initially. For each job $i_n = \gamma[n]$, $n \in \{1,2,\ldots, u\}$, we test it at all possible positions of $\lambda$, and insert $i_n$ at the position of λ resulting in the lowest $CM_{max}$. Finally, a complete job permutation λ is obtained.

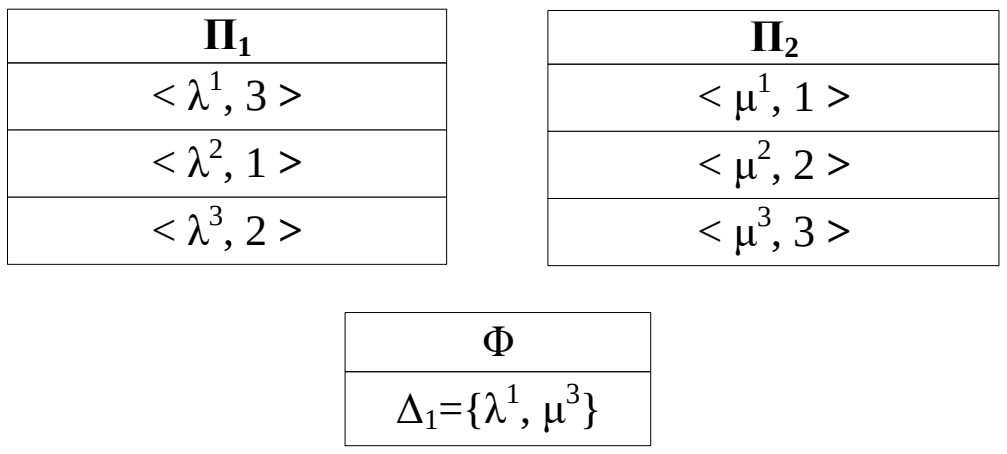

Fig. 7. *m*CCEA framework where *PS* = 3 and *AS* = 1.

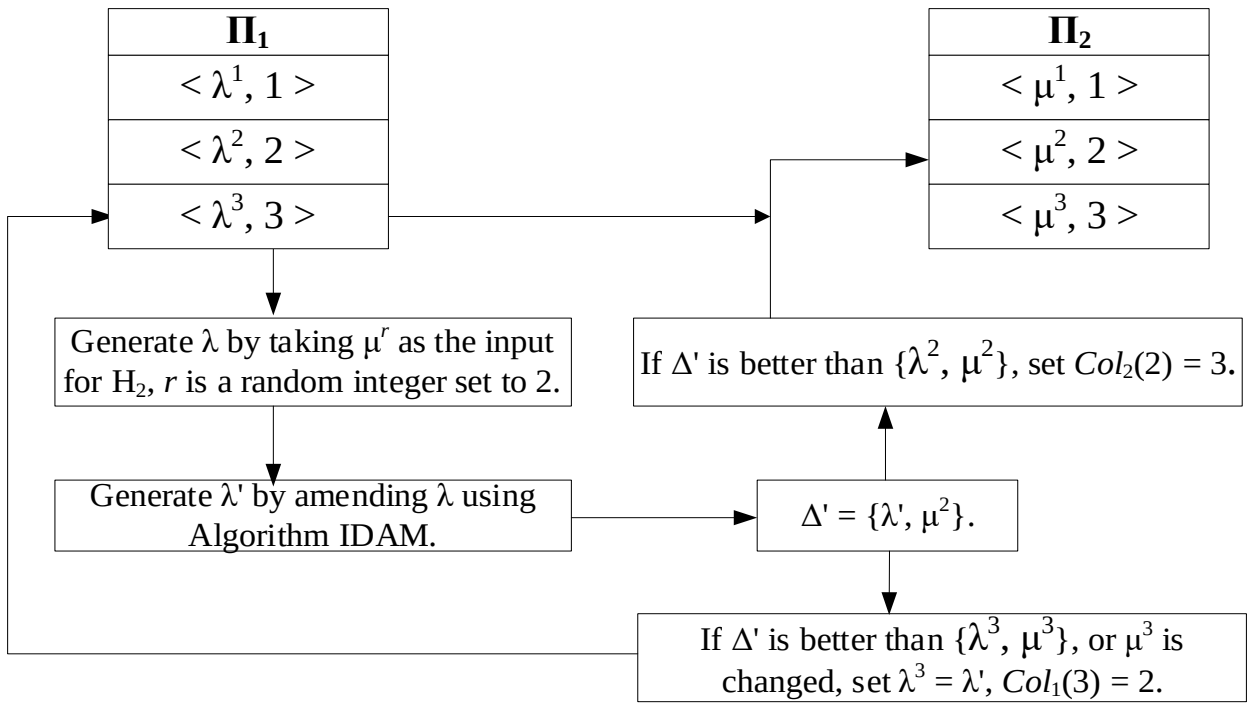

Fig. 8. An example of the evolution process for λ sequence.

Based on operators $H_1$ and $H_2$, we propose four initialization methods $L_1 - L_4$.

$L_1$: Generate a job permutation λ by sorting all jobs in descending order based on the sum of their processing time. Then, take λ as the input for $H_1$, obtain a factory permutation μ and an initial individual Δ = {λ, μ}.

$L_2$: Generate a permutation σ by sorting all products in descending order of the sum of their processing time. For each product $q \in \sigma$, sort jobs in $AP_q$ in descending order of the sum of their processing time, and obtain a sequence $S_q$. Finally, a job permutation λ is generated by sorting sequences $S_q$ according to σ. Taking λ as the input for $H_1$, we obtain a factory permutation μ and an individual Δ = {λ, μ}.

$L_3$: Take a random job permutation λ as the input for $H_1$, obtain a factory permutation μ and an initial individual Δ = {λ, μ}.

$L_4$: Take a random factory permutation μ as the input for $H_2$, obtain a job permutation λ and an initial individual Δ = {λ, μ}.

2) *Initialization for $\Pi_1$, $\Pi_2$ and $\Phi$*

We generate the first four individuals using the four heuristic methods $L_1$–$L_4$. The remaining *PS*–4 individuals are generated randomly. Algorithm IDAM is used to ensure the deadlock-freeness of all job permutations. The obtained individuals are denoted by $\Delta^n = \{\lambda^n, \mu^n\}$, $n \in \{1,2,\ldots, PS\}$.

Initially, for each $n = 1,2,\ldots, PS$, let $col_1[n] = col_2[n] = n$. Then $\Pi_1$ and $\Pi_2$ store all $<\lambda^n, col_1[n]>$ and $<\mu^n, col_2[n]>$, respectively. The top $AS = ep\% \times PS$ individuals with the shortest system makespan are selected to construct EAR.

*C. Global evolution for HCCE*

Note that $\Pi_1$ and $\Pi_2$ evolve with different problem-specific global explorers.

For each $<\lambda^n, col_1[n]>$ in $\Pi_1$, $n = 1,2,\ldots, PS$, let *r* be a random integer between 1 and *PS*. We obtain a new permutation λ by taking $\mu^r$ in $\Pi_2$ as the input of operator $H_2$. By Algorithm IDAM, we convert λ to deadlock-free λ′. If the individual $\{\lambda', \mu^r\}$ is better than $\{\lambda^n, \mu^{col1[n]}\}$, or $\mu^{col1[n]}$ in $\Pi_2$ has been changed, we set $\lambda^n = \lambda'$ and $col_1[n] = r$. Meanwhile, if $\{\lambda', \mu^r\}$ is better than $\{\lambda^{col2[r]}, \mu^r\}$, we set $col_2[r] = n$. An example of above evolution process is illustrated in Fig. 8.

Similarly, for each entity $<\mu^n, col_2[n]> \in \Pi_2$, let *r* be a random integer between 1 and *PS*. A new permutation μ is generated by taking $\lambda^r$ in $\Pi_1$ as the input for operator $H_1$. If the new individual $\{\lambda^r, \mu\}$ is better than $\{\lambda^{col2[n]}, \mu^n\}$, or $\lambda^{col2[n]}$ in $\Pi_1$ has been changed, we set $\mu^n = \mu$ and $col_2[n] = r$. Also, if $\{\lambda^r, \mu\}$ is better than $\{\lambda^r, \mu^{col1[r]}\}$, we set $col_1[r] = n$.

*D. Information transfer mechanism*

To improve the diversity and the quality of individuals, we propose an information transfer mechanism (ITM), transferring the superior coding among subpopulations $\Pi_1$, $\Pi_2$, and EAR Φ.

Denote $<\lambda^b, col_1[b]>$ and $<\lambda^w, col_1[w]>$ as the *best* and *worst* entities in $\Pi_1$ with the shortest and largest system makespan, respectively. Similarly, the best and worst entities in $\Pi_2$ is denoted by $<\mu^b, col_2[b]>$ and $<\mu^w, col_2[w]>$, respectively.

First, let $r_1$ and $r_2$ be two random integers between 1 and *PS*. Then for entities $<\lambda^{r1}, col_1[r_1]> \in \Pi_1$ and $<\mu^{r2}, col_2[r_2]> \in \Pi_2$, we replace $\lambda^{r1}$ with $\lambda^{col2[b]}$ and replace $\mu^{r2}$ with $\mu^{col1[b]}$.

Next, let $r_3$ and $r_4$ be two random integers between 1 and *AS*. The elite individuals $\Delta_e^{r3} = \{\lambda_e^{r3}, \mu_e^{r3}\}$ and $\Delta_e^{r4} = \{\lambda_e^{r4}, \mu_e^{r4}\}$ are replaced with $\{\lambda^b, \mu^{col1[b]}\}$ and $\{\lambda^{col2[b]}, \mu^b\}$, respectively. On the other hand, the permutations $\lambda^w$ and $\mu^w$ in the worst entities are replaced by $\lambda_e^{r3}$ and $\mu_e^{r4}$, respectively.

*E. Local search performed on EAR*

To increase the convergence ability of HCCE, four local search operators $LS_1$–$LS_4$ are executed sequentially on each elite individual $\Delta_e^n = \{\lambda_e^n, \mu_e^n\}$ in EAR Φ, where $n = 1,2,\ldots, AS$. Their details are as follows.

$LS_1$*:* Let $\lambda = \lambda_e^n$. For a product *q*, remove all jobs of $AP_q$ from λ and denote the remaining sequence as $\lambda_q$. Then we insert jobs of $AP_q$ sequentially into $\lambda_q$, leading to the lowest system makespan. Meanwhile, assign all jobs in $AP_q$ to the factory with the lowest system makespan. Continue this process for all products. If the generated new individual Δ is better than the original $\Delta_e^n$, replace it with Δ.

*Remark 4:* Identify $q_t$ as the last product that is not assembled continuously, and the factory assigned to the last processing job of, is called *critical-factory*. Meanwhile, the factory with the maximum start time for its first processing job is called *min-factory*. For example, consider the scheduling result in Fig. 6, the last product that is not assembled continuously is $q_1$, thus, $q_t = q_1$. The last processing job for $q_1$ is $i_3$, which is assigned to factory $c_1$. Thus, the *critical-factory* is $c_1$. Additionally, the *min-factory* is $c_2$.

$LS_2$: Let Γ be the set of jobs processed in the critical-factory. Remove all jobs in Γ from $\lambda_e^n$, and denote the remaining permutation as $\lambda_q$. Insert each job of Γ into the certain position of $\lambda_q$ that leads to the lowest system makespan. Finally, obtain a complete job permutation λ. Use Algorithm IDAM to convert λ to a deadlock-free λ′. If the generated new individual $\Delta = \{\lambda', \mu_e^n\}$ is better than $\Delta_e^n$, replace it with Δ.

$LS_3$: For each job processed by the critical-factory, we change its assigned factory to min-factory, and obtain a new factory permutation μ. If the generated new individual $\Delta = \{\lambda_e^n, \mu\}$ is better than $\Delta_e^n$, replace it with Δ.

$LS_4$: Denote $\Delta_{best} = \{\lambda_{best}, \mu_{best}\}$ as the best individual in Φ. We first randomly remove $d = u \times cd\%$ jobs from $\lambda_e^n$, and let Γ collect all these removed jobs. Then, a new job sequence λ is generated from $\lambda_e^n$ by arranging the jobs of Γ according to their order in $\lambda_{best}$. The assigned factory for each job in Γ is changed according to $\mu_{best}$, we obtain a new factory sequence μ from $\mu_e^n$. Use Algorithm IDAM to convert λ to deadlock-free λ′. If the generated new individual $\Delta = \{\lambda', \mu\}$ is better than the original one, replace it with Δ.

### *F. Restart*

To avoid the local optimum, we propose an reinitialization strategy to restart the populations for remaining the diversity. Specifically, if the best individual $\Delta_{best} = \{\lambda_{best}, \mu_{best}\}$ of EAR Φ has not been improved in a predetermined number α of consecutive generations, we restart the subpopulations $\Pi_1$ and $\Pi_2$ according to the initialization method proposed in section IV. B.

### *G. Overall HCCE algorithm for DAFSP*

The overall HCCE algorithm for DAFSP is outlined as follows.

**Algorithm 5** HCCE algorithm.

```
Set parameters AS, PS, and α.
Construct APP model for DAFSP.
Initialization for Π1, Π2 and Φ.
While termination is not satisfied do
    Perform global evolution for each entity in Π1.
    Perform global evolution for each entity in Π2.
    Exchange information among Π1, Π2, and Φ through
    ITM model.
    Perform local search for each individual in Φ.
    If the best individual Δbest ∈Φ is not improved in α
    consecutive generations
        Initialization for Π1 and Π2.
    end
end
Output the Δbest.
```

## V. COMPUTATION EXPERIMENTS

In this section, we conduct extensive experiments to evaluate the performance of proposed HCCE algorithm in solving various deadlock-prone DAFSPs. First, the parameter values of HCCE are calibrated via a design of experiment method. The components of HCCE are validated by three of its variant algorithms. Finally, we compare HCCE with three other state-of-the-art algorithms. The experiments are run on a Windows 10 operating system with an Intel Core i9-10900K CPU @ 3.70 GHz and 64.0 GB of RAM by using MATLAB programming language.

### *A. Experimental settings*

Since the DAFSP has not been dealt with before, according to the benchmark design idea [1], we generate three types of testing instances: *small*, *medium*, and *large*. The scale parameters *u*, *f*, *m*, and *l* for each instance type are listed in Table III, where *u*, *f*, *m*, and *l* are the numbers of jobs, factories, machines, and products, respectively. There are a total of 108 combinations of $u \times f \times m \times l$. For each combination, three cases are presented in which the processing times for jobs and products are randomly generated within [1, 99] and [1, 50], respectively, following a uniform distribution. Therefore, we consider 324 = 108 × 3 instances.

Table III
PARAMETER SIZES FOR EACH INSTANCE TYPE

| Instance type | $u$ | $f$ | $m$ | $l$ | $u \times f \times m \times l$ |
|---|---|---|---|---|---|
| Small | 10, 16, 24 | 2, 3 | 2, 4, 6 | 2, 4 | 3 × 2 × 3 × 2 = 36 |
| Medium | 30, 40, 50 | 4, 6 | 8, 10, 12 | 6, 8 | 3 × 2 × 3 × 2 = 36 |
| Large | 80, 100, 120 | 8, 10 | 16, 18, 20 | 10, 16 | 3 × 2 × 3 × 2 = 36 |

Table IV
ORTHOGONAL ARRAY $L_{16}(4^4)$ AND RESPONSE VARIABLES

| Trial | Factor Level | | | | Response Value | | |
|---|---|---|---|---|---|---|---|
| | *PS* | *ep* | α | *cd* | Small | Medium | Large |
| 1 | 1 | 1 | 1 | 1 | 445.5 | 1175.5 | 1769.7 |
| 2 | 1 | 2 | 2 | 2 | 449.1 | 1175.9 | 1775.4 |
| 3 | 1 | 3 | 3 | 3 | 445.7 | 1172.3 | 1777.2 |
| 4 | 1 | 4 | 4 | 4 | 444.1 | 1173.9 | 1773.9 |
| 5 | 2 | 1 | 2 | 3 | 441.4 | 1172.2 | 1775.1 |
| 6 | 2 | 2 | 1 | 4 | 446.4 | 1172.4 | 1776.5 |
| 7 | 2 | 3 | 4 | 1 | 446.6 | 1167.1 | 1784.3 |
| 8 | 2 | 4 | 3 | 2 | 445.6 | 1168.9 | 1779.7 |
| 9 | 3 | 1 | 3 | 4 | 444 | 1171.3 | 1775 |
| 10 | 3 | 2 | 4 | 3 | 445.6 | 1172.4 | 1771.8 |
| 11 | 3 | 3 | 1 | 2 | 448.4 | 1173.7 | 1774.7 |
| 12 | 3 | 4 | 2 | 1 | 442.4 | 1174.2 | 1775.8 |
| 13 | 4 | 1 | 4 | 2 | 443.9 | 1172.2 | 1774.7 |
| 14 | 4 | 2 | 3 | 1 | 442 | 1167.8 | 1775.7 |
| 15 | 4 | 3 | 2 | 4 | 446.1 | 1169.2 | 1776.1 |
| 16 | 4 | 4 | 1 | 3 | 446.4 | 1172.2 | 1775.7 |

Table V
STATISTICAL ANALYSIS AND SELECTED PARAMETER VALUE

| Factor level | | *PS* | *ep* | *α* | *cd* |
|---|---|---|---|---|---|
| Small | 1 | 446.1 | **443.7** | 446.675 | **444.125** |
| | 2 | 445 | 445.775 | 444.75 | 446.75 |
| | 3 | 445.1 | 446.7 | **444.325** | 444.775 |
| | 4 | **444.6** | 444.625 | 445.05 | 445.15 |
| | Delta | 1.5 | 3 | 2.35 | 2.625 |
| | Rank | 4 | 1 | 3 | 2 |
| | SPV | 50 | 0.2 | 21 | 0.1 |
| Medium | 1 | 1174.4 | 1172.8 | 1173.45 | **1171.15** |
| | 2 | **1170.15** | 1172.125 | 1172.875 | 1172.675 |
| | 3 | 1172.9 | **1170.575** | **1170.075** | 1172.275 |
| | 4 | 1170.35 | 1172.3 | 1171.4 | 1171.7 |
| | Delta | 4.25 | 2.225 | 3.375 | 1.525 |
| | Rank | 1 | 3 | 2 | 4 |
| | SPV | 35 | 0.5 | 21 | 0.1 |
| Large | 1 | **1774.05** | **1773.625** | **1774.15** | 1776.375 |
| | 2 | 1778.9 | 1774.85 | 1775.6 | 1776.125 |
| | 3 | 1774.325 | 1778.075 | 1776.9 | **1774.95** |
| | 4 | 1775.55 | 1776.275 | 1776.175 | 1775.375 |
| | Delta | 4.85 | 4.45 | 2.75 | 1.425 |
| | Rank | 1 | 2 | 3 | 4 |
| | SPV | 25 | 0.2 | 3 | 0.7 |

For each instance, the assembly plan $AP_q$ for each product $q \in L$ is randomly generated. The capacity $\Psi$ of buffer $B_A$ is randomly generated with ranges of $[b, 1.5b]$ where $b = \max\{|AP_q|, q \in L\}$. The termination time for all algorithms is defined as $Time \times u \times f \times m \times l$ milliseconds, where *Time* takes three values 20, 50, 120 for large, medium, and small instances, respectively.

### *B. Parameter calibration*

Four relevant parameters {*PS*, *ep*, *cd*, *α*} in the HCCE algorithm require calibration, where *PS* denotes the number of individuals, *ep* represents the proportion of selected superior individuals in EAR Φ, *cd* is the proportion of jobs removed in the *destruction* process of $LS_4$, and *α* indicates the number of largest consecutive generations in the restart strategy.

The above parameter calibration is performed by following two steps: i) experimentally selecting different values for each parameter as its factor levels, and ii) determining the most suitable parameter values for different instance types using the traditional Taguchi design of experiment (DOE) method [39]. Each instance is tested ten times independently, and the average system makespan $CA_{max}$ serves as the performance metric. This process is illustrated as follows:

First, we test ten values for each parameter on a medium-type instance with $u \times f \times m \times l = 30 \times 4 \times 10 \times 6$, while the other parameters are fixed. Four values that lead to the minimum average $CA_{max}$ are selected as the factor levels.

Table VI
RESULTS OF HCCE AND COMPARED ALGORITHMS

| Instance group | | | $HCCE_1$ | | $HCCE_2$ | | $HCCE_3$ | | HCCE | |
|---|---|---|---|---|---|---|---|---|---|---|
| | | | *bRPD* | *aRPD* | *bRPD* | *aRPD* | *bRPD* | *aRPD* | *bRPD* | *aRPD* |
| S | *u* | 10 | 0.0175 | 0.0425 | 0.0166 | 0.0426 | 0.0054 | 0.0072 | **0.0000** | **0.0044** |
| | | 16 | 0.0327 | 0.0797 | 0.0354 | 0.0763 | 0.0041 | 0.0161 | **0.0016** | **0.0137** |
| | | 24 | 0.0413 | 0.0723 | 0.0382 | 0.0702 | 0.0064 | 0.0215 | **0.0013** | **0.0173** |
| | *f* | 2 | 0.0255 | 0.0513 | 0.0227 | 0.0526 | 0.0057 | 0.0133 | **0.0014** | **0.0099** |
| | | 3 | 0.0354 | 0.0784 | 0.0374 | 0.0735 | 0.0048 | 0.0165 | **0.0005** | **0.0137** |
| | *m* | 2 | 0.0231 | 0.053 | 0.0212 | 0.0541 | 0.0030 | 0.0106 | **0.0007** | **0.0088** |
| | | 4 | 0.0332 | 0.0759 | 0.0385 | 0.0702 | 0.0083 | 0.0199 | **0.0000** | **0.0135** |
| | | 6 | 0.0352 | 0.0656 | 0.0305 | 0.0648 | 0.0046 | 0.0143 | **0.0021** | **0.0131** |
| | *l* | 2 | 0.0294 | 0.0642 | 0.0294 | 0.0602 | 0.0042 | 0.0154 | **0.0014** | **0.0139** |
| | | 4 | 0.0315 | 0.0655 | 0.0307 | 0.0659 | 0.0063 | 0.0144 | **0.0005** | **0.0098** |
| | Avg | | 0.0305 | 0.0648 | 0.0301 | 0.0630 | 0.0053 | 0.0149 | **0.0010** | **0.0118** |
| M | *u* | 30 | 0.0491 | 0.075 | 0.0500 | 0.0725 | 0.0096 | 0.0239 | **0.0002** | **0.0135** |
| | | 40 | 0.0481 | 0.0725 | 0.0467 | 0.0712 | 0.0085 | 0.0211 | **0.0015** | **0.0134** |
| | | 50 | 0.047 | 0.0681 | 0.0456 | 0.0674 | 0.0120 | 0.0252 | **0.0006** | **0.0163** |
| | *f* | 4 | 0.045 | 0.0683 | 0.0484 | 0.0701 | 0.0113 | 0.0253 | **0.0005** | **0.0131** |
| | | 6 | 0.0511 | 0.0754 | 0.0465 | 0.0706 | 0.0087 | 0.0215 | **0.0010** | **0.0156** |
| | *m* | 8 | 0.0508 | 0.0758 | 0.0491 | 0.0720 | 0.0140 | 0.0254 | **0.0002** | **0.0150** |
| | | 10 | 0.0453 | 0.0731 | 0.0478 | 0.0725 | 0.0091 | 0.0240 | **0.0009** | **0.0164** |
| | | 12 | 0.0481 | 0.0666 | 0.0455 | 0.0666 | 0.0068 | 0.0208 | **0.0012** | **0.0117** |
| | *l* | 6 | 0.0458 | 0.0702 | 0.0458 | 0.0692 | 0.0098 | 0.0231 | **0.0009** | **0.0140** |
| | | 8 | 0.0504 | 0.0735 | 0.0491 | 0.0716 | 0.0102 | 0.0237 | **0.0006** | **0.0147** |
| | Avg | | 0.0481 | 0.0719 | 0.0475 | 0.0704 | 0.0100 | 0.0234 | **0.0008** | **0.0144** |
| L | *u* | 80 | 0.0436 | 0.0648 | 0.0354 | 0.0592 | 0.0088 | 0.0161 | **0.0000** | **0.0106** |
| | | 100 | 0.0352 | 0.0495 | 0.0301 | 0.0469 | 0.0046 | 0.0136 | **0.0007** | **0.0126** |
| | | 120 | 0.0277 | 0.0449 | 0.0282 | 0.0430 | 0.0060 | 0.0114 | **0.0000** | **0.0097** |
| | *f* | 8 | 0.0338 | 0.0512 | 0.0299 | 0.0481 | 0.0051 | 0.0128 | **0.0003** | **0.0102** |
| | | 10 | 0.0371 | 0.0549 | 0.0325 | 0.0512 | 0.0079 | 0.0146 | **0.0001** | **0.0117** |
| | *m* | 16 | 0.0409 | 0.0598 | 0.0381 | 0.0567 | 0.0082 | 0.0162 | **0.0003** | **0.0123** |
| | | 18 | 0.0353 | 0.0516 | 0.0326 | 0.0507 | 0.0078 | 0.0153 | **0.0003** | **0.0132** |
| | | 20 | 0.0302 | 0.0478 | 0.0230 | 0.0416 | 0.0034 | 0.0096 | **0.0001** | **0.0074** |
| | *l* | 10 | 0.0338 | 0.0509 | 0.0300 | 0.0490 | 0.0059 | 0.0128 | **0.0004** | **0.0108** |
| | | 16 | 0.0372 | 0.0552 | 0.0324 | 0.0504 | 0.0070 | 0.0146 | **0.0001** | **0.0111** |
| | Avg | | 0.0355 | 0.0531 | 0.0312 | 0.0497 | 0.0065 | 0.0137 | **0.0002** | **0.0110** |

Specifically, the four levels chosen for *PS* are {25, 35, 40, 50}; for *ep*, {0.2, 0.3, 0.5, 0.6}; for *α*, {3, 6, 21, 24}; and for *cd*, {0.1, 0.4, 0.7, 0.8}.

Second, an orthogonal array $L_{16}(4^4)$ is established based on the different factor levels in Table IV. Since three instance types are designed for the DAFSP, DOE is performed three times for instances with $u \times f \times m \times l = 16 \times 3 \times 4 \times 4$, $40 \times 4 \times 10 \times 8$, and $80 \times 8 \times 18 \times 10$, respectively. During DOE execution, there are 16 parameter combinations, and HCCE is independently conducted 10 times for each combination. The average system makespan is used as the response value.

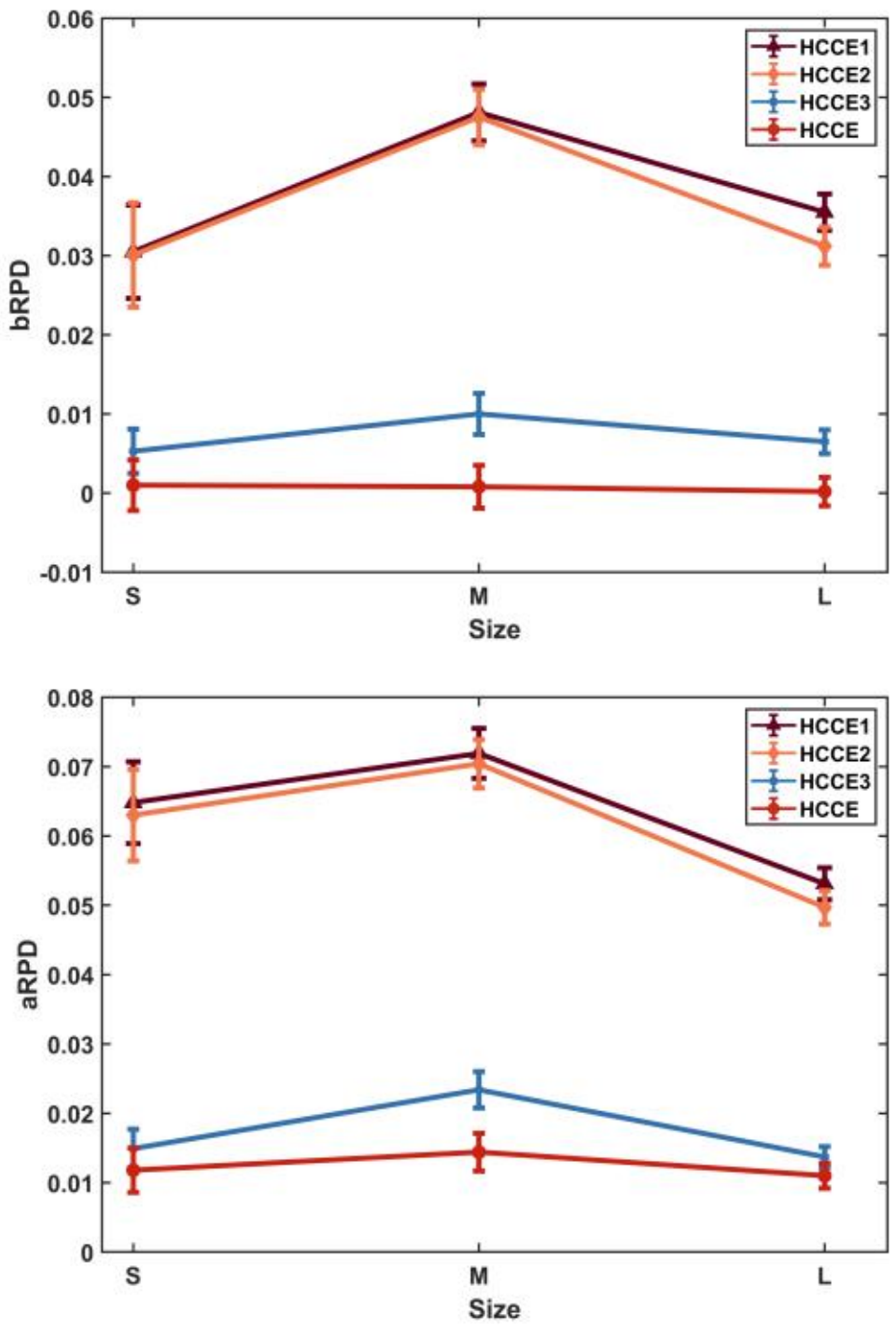

Fig. 9. Variation trend for compared algorithms.

Table VII
RESULTS OF THE FREIDMAN TEST

| | Instance group | Average rank | | | | Chi-Square | *P*-value |
|---|---|---|---|---|---|---|---|
| | | $HCCE_1$ | $HCCE_2$ | $HCCE_3$ | HCCE | | |
| *bRPD* | S | 3.65 | 3.35 | 2 | 1 | 27.545 | 0.0 |
| | M | 3.65 | 3.35 | 2 | 1 | 27.545 | 0.0 |
| | L | 3.90 | 3.10 | 2 | 1 | 28.920 | 0.0 |
| *aRPD* | S | 3.60 | 3.40 | 2 | 1 | 27.120 | 0.0 |
| | M | 3.85 | 3.15 | 2 | 1 | 28.758 | 0.0 |
| | L | 4 | 3 | 2 | 1 | 30.000 | 0.0 |

Table V presents a statistical analysis of the obtained results. In small-scale instances, *ep* is the most important factor, followed by *cd*, *α*, and *PS*. In medium-scale instances, *PS* is the most important factor, followed by *α*, *ep*, and *cd*. In large-scale instances, *PS* ranks as the most important factor, followed by *ep*, *α*, and *cd*. Therefore, the recommended parameter sets {*PS*, *ep*, *α*, *cd*} for small-scale, medium-scale, and large-scale instances are {50, 0.2, 21, 0.1}, {35, 0.5, 21, 0.1}, and {25, 0.2, 3, 0.7}, respectively.

### *C. Effectiveness of HCCE special components*

To ascertain the effectiveness of HCCE's components, we developed three variants of HCCE, denoted as $HCCE_1$–$HCCE_3$. $HCCE_1$ employs the basic CCEA-RS algorithm [40] with random initialization of individuals. $HCCE_2$ is $HCCE_1$ adding heuristic initialization. $HCCE_3$ is $HCCE_2$ incorporating the mCCEA framework. Finally, HCCE is $HCCE_3$ plus the local search method.

Table VIII
RESULTS OF HCCE AND COMPARED ALGORITHMS

| Instance group | | | HHMA | | EDMBO | | PBIGA | | HCCE | |
|---|---|---|---|---|---|---|---|---|---|---|
| | | | *bRPD* | *aRPD* | *bRPD* | *aRPD* | *bRPD* | *aRPD* | *bRPD* | *aRPD* |
| S | *u* | 10 | 0.0875 | 0.1035 | 0.0375 | 0.0877 | 0.0304 | 0.0431 | **0.0010** | **0.0081** |
| | | 16 | 0.1172 | 0.1299 | 0.0480 | 0.0922 | 0.0374 | 0.0524 | **0.0003** | **0.0102** |
| | | 24 | 0.1338 | 0.1552 | 0.0330 | 0.0687 | 0.0490 | 0.0641 | **0.0007** | **0.0160** |
| | *f* | 2 | 0.1091 | 0.1261 | 0.0290 | 0.0660 | 0.0338 | 0.0453 | **0.0005** | **0.0096** |
| | | 3 | 0.1166 | 0.1330 | 0.0500 | 0.0998 | 0.0440 | 0.0612 | **0.0008** | **0.0132** |
| | *m* | 2 | 0.1142 | 0.1276 | 0.0351 | 0.0781 | 0.0339 | 0.0490 | **0.0007** | **0.0116** |
| | | 4 | 0.1185 | 0.1400 | 0.0423 | 0.0868 | 0.0385 | 0.0513 | **0.0011** | **0.0123** |
| | | 6 | 0.1058 | 0.1210 | 0.0412 | 0.0837 | 0.0443 | 0.0593 | **0.0002** | **0.0104** |
| | *l* | 2 | 0.1305 | 0.1439 | 0.0323 | 0.0693 | 0.0418 | 0.0566 | **0.0009** | **0.0113** |
| | | 4 | 0.0952 | 0.1152 | 0.0468 | 0.0983 | 0.0360 | 0.0498 | **0.0005** | **0.0115** |
| | Avg | | 0.1128 | 0.1295 | 0.0395 | 0.0831 | 0.0389 | 0.0532 | **0.0007** | **0.0114** |
| M | *u* | 30 | 0.0413 | 0.0573 | 0.0609 | 0.1271 | 0.0186 | 0.0320 | **0.0006** | **0.0152** |
| | | 40 | 0.0462 | 0.0621 | 0.0487 | 0.1054 | 0.0191 | 0.0305 | **0.0003** | **0.0146** |
| | | 50 | 0.0429 | 0.0564 | 0.0461 | 0.0998 | 0.0144 | 0.0250 | **0.0017** | **0.0156** |
| | *f* | 4 | 0.0494 | 0.0655 | 0.0445 | 0.1006 | 0.0205 | 0.0325 | **0.0008** | **0.0150** |
| | | 6 | 0.0376 | 0.0517 | 0.0594 | 0.1210 | 0.0143 | 0.0258 | **0.0010** | **0.0153** |
| | *m* | 8 | 0.0477 | 0.0630 | 0.0546 | 0.1166 | 0.0211 | 0.0339 | **0.0012** | **0.0161** |
| | | 10 | 0.0424 | 0.0579 | 0.0473 | 0.1060 | 0.0141 | 0.0270 | **0.0012** | **0.0150** |
| | | 12 | 0.0404 | 0.0549 | 0.0539 | 0.1097 | 0.0169 | 0.0266 | **0.0002** | **0.0144** |
| | *l* | 6 | 0.0421 | 0.0565 | 0.0483 | 0.1040 | 0.0196 | 0.0310 | **0.0006** | **0.0154** |
| | | 8 | 0.0449 | 0.0607 | 0.0555 | 0.1178 | 0.0152 | 0.0273 | **0.0012** | **0.0149** |
| | Avg | | 0.0435 | 0.0586 | 0.0519 | 0.1108 | 0.0174 | 0.0292 | **0.0009** | **0.0152** |
| L | *u* | 80 | 0.0275 | 0.0422 | 0.0635 | 0.1278 | 0.0065 | 0.0135 | **0.0004** | **0.0102** |
| | | 100 | 0.0244 | 0.0366 | 0.0489 | 0.0949 | 0.0063 | 0.0120 | **0.0012** | **0.0109** |
| | | 120 | 0.0281 | 0.0399 | 0.0603 | 0.0985 | 0.0111 | 0.0163 | **0.0015** | **0.0109** |
| | *f* | 8 | 0.0259 | 0.0394 | 0.0488 | 0.0949 | 0.0069 | 0.0132 | **0.0008** | **0.0105** |
| | | 10 | 0.0274 | 0.0398 | 0.0663 | 0.1192 | 0.0090 | 0.0147 | **0.0013** | **0.0108** |
| | *m* | 16 | 0.0254 | 0.0390 | 0.0612 | 0.1130 | 0.0071 | 0.0135 | **0.0019** | **0.0123** |
| | | 18 | 0.0244 | 0.0368 | 0.0549 | 0.1025 | 0.0062 | 0.0118 | **0.0007** | **0.0095** |
| | | 20 | 0.0302 | 0.0429 | 0.0566 | 0.1056 | 0.0106 | 0.0166 | **0.0005** | **0.0102** |
| | *l* | 10 | 0.0224 | 0.0343 | 0.0547 | 0.0985 | 0.0074 | 0.0141 | **0.0011** | **0.0102** |
| | | 16 | 0.0310 | 0.0448 | 0.0604 | 0.1153 | 0.0086 | 0.0138 | **0.0010** | **0.0112** |
| | Avg | | 0.0267 | 0.0396 | 0.0576 | 0.1070 | 0.0080 | 0.0140 | **0.0010** | **0.0107** |

We compare HCCE with $HCCE_1$-$HCCE_3$ based on the merit called relative percentage deviation (*RPD*), which is defined as follows:

$$RPD = (CA_{max} - CA_{max}^{best}) / CA_{max}^{best} \quad (10)$$

where $CA_{max}$ is the system makespan of the current algorithm for a test instance, and $CA_{max}^{best}$ is the lowest system makespan obtained by all algorithms for the same instance. To reduce randomness, after running each algorithm 10 times independently, we record the best *RPD* (*bRPD*) and average *RPD* (*aRPD*).

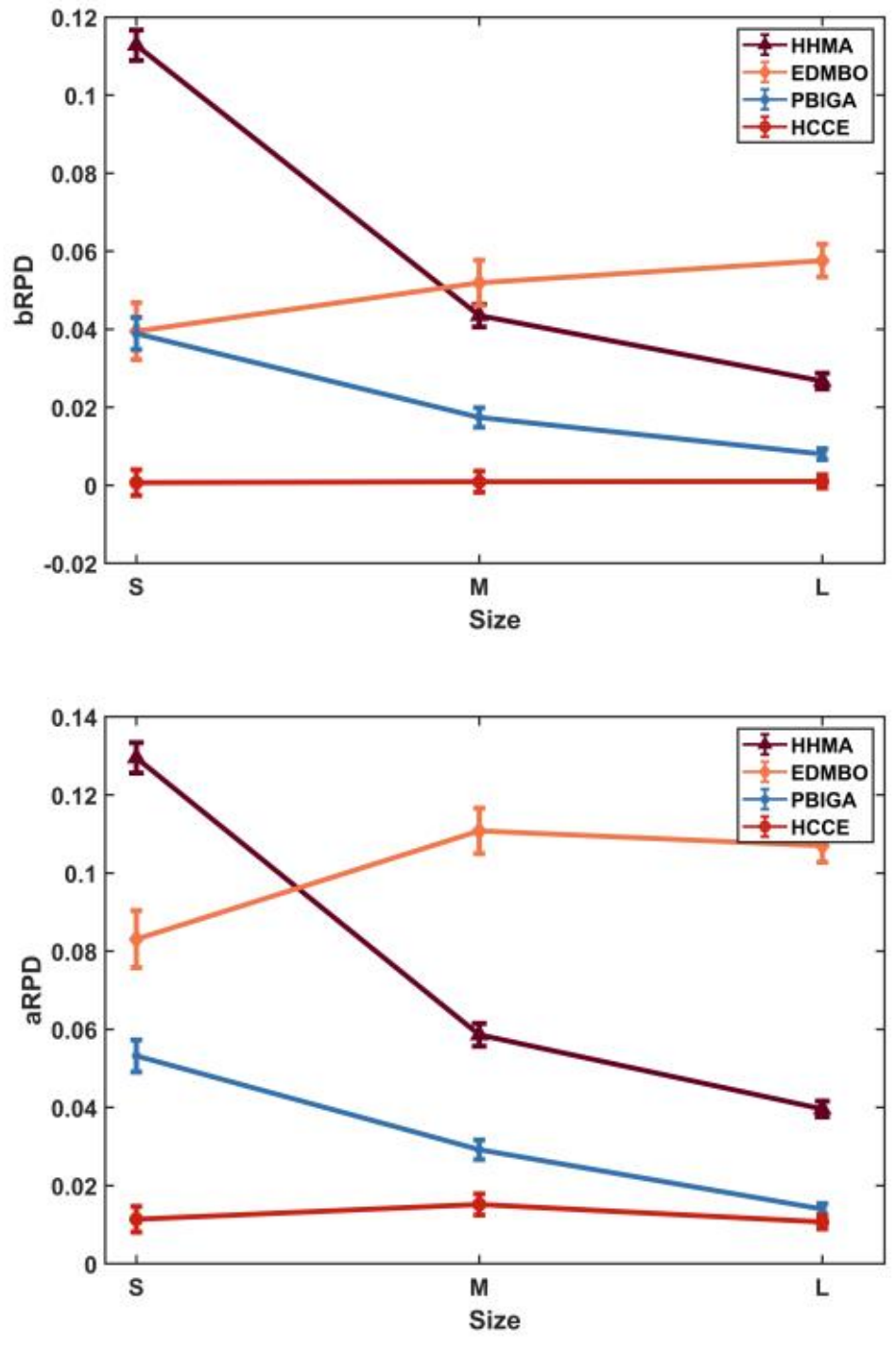

Fig. 10. Variation trend for compared algorithms.

Table IX
RESULTS OF THE FREIDMAN TEST

| | Instance group | Average rank | | | | Chi-Square | P-value |
|---|---|---|---|---|---|---|---|
| | | HHMA | EDMBO | PBIGA | HCCE | | |
| *bRPD* | S | 4 | 2.6 | 2.4 | 1 | 27.120 | 0.0 |
| | M | 3.1 | 3.9 | 2.0 | 1 | 28.920 | 0.0 |
| | L | 3 | 4 | 2 | 1 | 30.000 | 0.0 |
| *aRPD* | S | 4 | 3 | 2 | 1 | 30.000 | 0.0 |
| | M | 3 | 4 | 2 | 1 | 30.000 | 0.0 |
| | L | 4 | 3 | 2 | 1 | 30.000 | 0.0 |

Comparison results are summarized in Table VI by four levels of the number of jobs, factories, machines, and products. The best results for each category are marked in bold. The HCCE algorithm outperforms other variants in all categories. Fig. 9 plots the trend variations of *bRPD* and *aRPD* for $HCCE_1$–$HCCE_3$, and HCCE under each instance type. It is observed that: 1) $HCCE_1$ has the worst performance in all instances, 2) $HCCE_2$ shows slight improvement over $HCCE_1$, suggesting that the initialization strategy moderately enhances algorithm performance, 3) $HCCE_3$ significantly outperforms $HCCE_2$, 4) HCCE is better than $HCCE_3$, indicating the effectiveness of the local search method, and 5) the most substantial improvement is observed with the transition from $HCCE_2$ to $HCCE_3$. That means the mCCEA framework plays a vital role in enhancing HCCE performance.

To enhance the credibility of our results, a statistical analysis is performed to validate the comparison algorithms within a 95% confidence interval and to detect significant differences, utilizing the Friedman test [41]. The results of the Freidman test are summarized in Table VII by instance size, showing all *P*-values are 0 and the HCCE algorithm consistently achieved the lowest average rank, indicating the HCCE algorithm outperforms other variants in all instances.

### *D. Comparison with state-of-the-art algorithms*

To evaluate the effectiveness of the HCCE algorithm, we compare it with three state-of-the-art algorithms: Hyper Heuristic-based Memetic Algorithm (HHMA) [7], Effective Discrete Monarch Butterfly Optimization (EDMBO) algorithm [8], and Population-Based Iterated Greedy Algorithm (PBIGA) [6]. These algorithms and HCCE are applied to the model and experimental settings in this article. The instances are run 10 times independently and the corresponding *bRPD* and *aRPD* are obtained.

Comparison results are summarized in Table VIII by three different instance types. The best results in each type are marked in bold. The HCCE algorithm outperforms other compared algorithms for all instance types. Fig. 10 illustrates the trend variations of *bRPD* and *aRPD* for all algorithms. It shows that: 1) in small-type instances, HCCE significantly outperforms the other algorithms, while HMMA exhibits the worst performance, and EDMBO and PBIGA show similar performance, 2) in other instance types, HCCE continues to perform best, but EDMBO is the worst, and the performance of PBIGA gradually approaches that of HCCE, particularly in large-type instances.

A statistical analysis is conducted to verify the comparison results with a 95% confidence interval, identifying significant differences among comparison algorithms. The Freidman test results are summarized in Table IX by instance size, showing all *P*-values are 0 and HCCE has the lowest average rank among all algorithms. Experimental results confirm that the HCCE algorithm outperforms the other comparison algorithms in all instances.

## VI. CONCLUSION

Deadlock is a crucial problem for DAFSP with limited buffers, and to date no one has well resolved this issue. For the first time, this paper utilizes Petri nets to model deadlocks in DAFSP and proposes an amending method, namely IDAM, to ensure the feasibility or deadlock-freeness of the solutions for DAFSP. This method is integrated into the proposed HCCE algorithm, which has several special designs. First, an mCCEA framework that includes two subpopulations and EAR is proposed. To improve the quality and diversity of individuals, an ITM is used to exchange information among subpopulations, and two problem-specific operators are introduced for initialization and global search. Further, four local search operators are sequentially applied to each individual in the EAR to improve the searchability. Finally, the effectiveness of the components of HCCE is validated by comparing it with its three variants. The results demonstrate that HCCE outperforms the variants and the mCCEA framework is most key factor for enhancing the performance. Meanwhile, the experiment results show that HCCE outperforms three state-of-the-art comparison algorithms, including HHMA [7], EDMBO [8], and PBIGA [6].

Future work should explore DAFSP with multiple assembly factories and dynamic scheduling scenarios, such as urgent product insertions. Additionally, more strategies such as reinforcement learning and acceleration method should be considered.

## Appendix

A PN is a three-tuple $N = (P, T, F)$, where $P$ and $T$ are finite and disjoint sets. $P = \{p_1,p_2,\ldots,p_r\}$ is the set of places, where $r$ is the number of places. $T = \{t_1,t_2,\ldots,t_d\}$ is the set of transitions, where $d$ is the number of transitions. $F \subseteq (P \times T) \cup (T \times P)$ is the set of directed arcs, where $P \times T$ denotes the directed arcs from $P$ to $T$ and $T \times P$ denotes the directed arcs from $T$ to $P$. Places, transitions, and directed arcs constitute a directed graph, where $P \cup T$ is the set of vertices and $F$ is the set of arcs of the directed graph. $P$, $T$, and $F$ are represented by circles, rectangles, and arcs with arrows respectively.

Given a node $x \in P \cup T$, the preset of $x$ is defined as ${}^\bullet x = \{y \in P \cup T \mid (y, x) \in F\}$, and the postset of $x$ is defined as $x^\bullet = \{y \in P \cup T \mid (x, y) \in F\}$. These notations can be extended to a set, for example, let $X \in P \cup T$ and then ${}^\bullet X = \cup_{x \in X} {}^\bullet x$ and $X^\bullet = \cup_{x \in X} x^\bullet$.

A marking or state of $N$ is a mapping $M : P \rightarrow \mathbb{Z}$, where $\mathbb{Z} = \{0,1,2,\ldots\}$. Given a place $p \in P$ and a marking $M$, $M(p)$ denotes the number of tokens in $p$ at $M$. Let $S \subseteq P$ be a set of places; the sum of tokens in all places of $S$ at $M$ is denoted by $M(S)$, i.e., $M(S) = \sum_{p \in S} M(p)$. A PN $N$ with an initial marking $M_0$ is called a marked PN, denoted as $(N, M_0)$.

A transition $t \in T$ is enabled at a marking $M$, denoted by $M[t>$, if $\forall p \in {}^\bullet t$, $M(p) > 0$. An enabled transition $t$ at $M$ can result in a new reachable marking $M'$, denoted by $M[t> M'$, where $M'(p) = M(p) - 1$, $\forall p \in {}^\bullet t \setminus t^\bullet$, $M'(p) = M(p) + 1$, $\forall p \in t^\bullet \setminus {}^\bullet t$, otherwise, $M'(p) = M(p)$. A sequence of transitions $\alpha = t_1 t_2 \ldots t_k$ is *feasible* from a marking $M$ if there exists $M_i[t_i > M_{i+1}$, $i = 1,2,\ldots,k$, where $M_1 = M$. We state that $M_i$ is a reachable marking from $M$.